\documentclass[twocolumn,showpacs,aps,prd,nobibnotes,nofootinbib,floatfix]{revtex4-2}

\usepackage{amsmath}
\usepackage{graphicx,subfigure}
\usepackage[usenames]{color}
\usepackage[normalem]{ulem}
\usepackage{textcomp}
\usepackage{gensymb}
\usepackage{xspace}
\usepackage{verbatim}
\usepackage{soul}
\usepackage[colorlinks=true]{hyperref}
\hypersetup{
  citecolor  = blue
}

\begin{document}

\title{Weak cosmic censorship conjecture is not violated for a rotating linear dilaton black hole }

\author{Fei Qu$^a$$^b$}
\author{Si-Jiang Yang$^a$$^b$}
\author{Zhi Wang$^a$}
\author{Ji-Rong Ren$^a$$^b$$^c$}
\email{renjr@lzu.edu.cn, corresponding author}

\affiliation{$^{a}$Lanzhou Center for Theoretical Physics, Lanzhou University, Lanzhou, Gansu 730000, China \\
$^{b}$Institute of Theoretical Physics $\&$ Research Center of Gravitation, Lanzhou University, Lanzhou 730000, China\\
$^{c}$Key Laboratory for Magnetism and Magnetic of the Ministry of Education, Lanzhou University, Lanzhou 730000, China}

\begin{abstract}
This paper investigates the validity of the weak cosmic censorship conjecture (WCCC) for a rotating linear dilaton black hole in the Einstein-Maxwell-Dilaton-Axion (EMDA) theory. We firstly compare the definition of conserved charge obtained by several ways for this black hole solution. The variation of the corresponding conserved charge is proven to be consistent among these methods. We then test if WCCC is preserved by classically perturbation of particle. It turns out to be the same up to the same order approximation as the result given in the new version of gedanken experiment recently proposed by Wald. Even in the linear dilaton case, the nearly extremal black hole can be overspun, while the extremal one cannot be overspun up to the first order approximation. When we incorporate the second order modification into consideration, the Iyer-Wald method shows that even for the nearly extremal black hole, the WCCC is well protected. These results imply that weak cosmic censorship conjecture is still valid using the ingoing test particle method up to second order modification.
\end{abstract}

\maketitle

\section{Introduction}

Gravitational collapse inevitably leads to space-time singularity, which indicates the failure of the predictability of the theory. To protect the predictability of classical gravitational theory, Penrose proposed the weak cosmic censorship conjecture, which states that naked singularity cannot be formed by gravitational collapse~\cite{Penr69}. Though more than fifty years has passed while a general proof of the conjecture is still beyond reach, the conjecture has play an important role in black hole physics. Over the past fifty years, many ways have been proposed to test the conjecture~\cite{East19,FiKT16,FiKT17,CrSa17,NaQV16,HSSLW20,SoHW20}. One way of checking the conjecture is to consider whether we can destroy the event horizon to form a naked singularity.

Pioneer work to consider the destruction of a black hole event horizon was envisaged by Wald, in whose gedanken experiment a test particle with large charge or large angular momentum was dropped into an extremal Kerr-Newman black hole. The result suggests that particles causing the destruction of event horizon will not be captured by a black hole~\cite{Wald74}. While, later work of Hubeny shows that near-extremal charged black holes can be overcharged by test particle~\cite{Hube99}, the results are supported by the following work of Jacobson and Sotiriou for Kerr black hole~\cite{JaSo09}. By carefully choosing the parameters of the particle, there are some possible counter-examples that the event horizon of near-extremal black holes can be destroyed~\cite{YCWWL20,YWCYW20,FYTYL20,LiWL19}. But when self-force and radiation effects are taken into account, the above counter-example for weak cosmic censorship conjecture might be restored~\cite{BaCK10,BaCK11,ZVPH13,CoBa15,CBSv15}.

Recently, Sorce and Wald proposed a new version of gedanken experiment by taking the second-order approximation of the perturbation that comes from the matter fields into account~\cite{SoWa17}. The new gedanken experiment is based on the Lagrangian method~\cite{Lee:1990nz,Iyer:1994ys,Iyer:1995kg,Wald:1999wa}, it naturally incorporates the self-force and backreaction effects. They showed that the Kerr-Newmann black hole cannot be overcharged or overspun up to second order perturbation. After that, in several cases with an asymptotically flat metric the validity of the conjecture has been confirmed~\cite{Jiang:2020mws,Jiang:2019soz,He:2019mqy,Shaymatov:2019pmn,Shaymatov:2018fmp,Ge:2017vun,Jiang:2019vww,Gao:2001ut,Jiang:2019ige,Wang:2020vpn,An:2017phb,Chen:2020hjm}.

Previous researches have not discussed the general cases with non-trivial background using this formulation~\cite{Chen:2019nhv,Wang:2019bml}. As for the general case, the black hole mechanics law in the Noether charge method may not directly correspond to the one using by other formulation. By comparison of the variation of the conserved charge, among the results given by Wald formulation, the results obtained by Hamiton-Jacobi methods and the results calculated from Komar integral minus background contribution, we find the above consideritions are consistent with each other~\cite{Brown:1992br}.

After that we test the WCCC by perturbing it with an ingoing particle. We find that the WCCC is valid for the extremal black hole, while not for the near extremal case. Then we review the derivation of the variational inequality. After that, by suitable constructing the perturbation process and evaluating variation equality on a certain region of the spacetime, we find it gives the same answer as the classical perturbation methods as mentioned up to first order approximation with the energy condition. If we further incorporate the second order modification into the case, we find the WCCC is well preserved. The similarity implies that if we consider perturbations of classical ingoing particle up to second order approximation, the WCCC might be well protected as well.

The rest of the paper is organised as follows. In Sec.~\ref{bh}, we give an overview for the linear dilaton solution from the EMDA theory.
In Sec.~\ref{komar}, we calculate the corresponding Komar-type integral for later convenience. In Sec.~\ref{CGE}, we discuss the classical version of the gedanken experiment. In Sec.~\ref{komar}, we review the Iyer-Wald formulation. We test the WCCC with the variational inequality in Sec.~\ref{W1} and Sec.~\ref{v1}. The last section is devoted to the conclusion.

\section{a rotating linear dilaton black hole IN EMDA THEORY}\label{bh}

In this section, we give a brief review concerned with a special black hole in the EMDA theory, then we will mention about its first law of thermodynamics.

The action for the EMDA is given by the action
\begin{eqnarray}
  S=&\nonumber \frac{1}{16 \pi} \int \sqrt{-g}d^{4} x [R-2 \partial_{\mu} \phi \partial^{\mu} \phi-\frac{1}{2} e^{4 \phi} \partial_{\mu} \kappa \partial^{\mu} \kappa\\
  &-e^{-2 \phi} F_{\mu \nu} F^{\mu \nu}-\kappa F_{\mu \nu} \star F^{\mu \nu}],
 \end{eqnarray}
 where $R$ is the usual Ricci scalar curvature, $\phi$ and $\kappa$ are the dilaton field and axion field, respectively. $\star$ is the usual Hodge dual operator on 4-d spacetime. $F_{\mu\nu}$ with its Hodge dual $\star F^{\mu\nu}$ corresponds to the  Maxwell field.

 It is well known that the lagrangian for the action is equivalent to the following formulation
 \begin{align}
 S=&\nonumber\int \sqrt{-g}d^{4} x [R-2 \partial_{\mu} \phi \partial^{\mu} \phi-\frac{1}{12} e^{-4 \phi} H_{\mu \nu \tau} H^{\mu \nu \tau}
 \\ &-\frac{1}{8} e^{-2 \phi} F_{\mu \nu} F^{\mu \nu}].
 \end{align}
 By ``equivalence", we mean that if we substitute
 \begin{equation}
 H^{\mu \nu \rho}=-\frac{1}{\sqrt{-g}} e^{4 \phi} \epsilon^{\mu \nu \rho \sigma} \partial_{\sigma} \kappa
 \end{equation}
 into the lagrangian and make a rescale, we can obtain the same equation of motion. One may seek more information about the equivalence of these two theory in \cite{GoulartSantos:2017dun,Matos:2009rp}. $ \epsilon^{\mu \nu \rho \sigma}$ is the total antisymmetric symbol. One asymptotically flat black hole solution has been discussed in \cite{Jiang:2019ige,Li:2010qh}.

 The stationary linear dilaton black hole solution  in spherical coordinate without nut charge is given
 as~\cite{Clement:2002mb}.
 \begin{equation} \label{metric1}
 d s^{2}=-\frac{\Gamma}{r_{0} r} d t^{2}+r_{0} r\left[\frac{d r^{2}}{\Gamma}+d \theta^{2}+\sin ^{2} \theta\left(d \varphi-\frac{a}{r_{0} r}dt\right)^{2}\right],
 \end{equation}
 with
 \begin{equation}\label{GAMMA}
 \Gamma=r^{2}-2 M r+a^{2}.
 \end{equation}
 The other fields are given by
\begin{equation}\label{fieldeq}
\begin{array}{l}
F=\frac{1}{\sqrt{2}}\left[\frac{r^{2}-a^{2} \cos ^{2} \theta}{r_{0} r^{2}} d r \wedge d t+a \sin 2 \theta d \theta \wedge\left(d \varphi-\frac{a}{r_{0} r} d t\right)\right], \\
e^{-2 \phi}=\frac{r_{0} r}{r^{2}+a^{2} \cos ^{2} \theta}, \\
\kappa=-\frac{r_{0} a \cos \theta}{r^{2}+a^{2} \cos ^{2} \theta},
\end{array}
\end{equation}
$F$, $\phi$ and $\kappa$ give the explicit expression of the electromagnetic field, dilaton field, and axion field, respectively.

 It is not difficult to recognise that  $t=const;r \to \infty$ these two conditions, give us a two sphere, which is so-called spatial infinity in spherical coordinate. The corresponding background metric regarding the spatial infinity metric is
 \begin{equation}\label{asympdia}
 d s^{2}=\frac{r}{r_{0}} d t^{2}-\frac{r_{0}}{r} d r^{2}+r_{0}r d \Omega^{2}.
 \end{equation}

 We may regard spatial infinity as the boundary of 3-Cauchy surface of the linear dilaton black hole. It is clear that different choice of $r_0$ will give us different background metric and different spatial infinity. Then the constancy in the variation process of $r_0$ is one of the most import feature of metric. So for the fixed boundary variation, the $r_0$ should be held as constant. In our case the   $r_0$  will have a constant correspondence with the electric charge. That is to say, the total electric charge over the manifold should be held invariant so as to have a well defined boundary to variate. This is why we call the metric a rotating linear dilaton black hole. The $M$ and $a$ are given as the parameters to determine the black hole properties.

 The event horizon is given by $\Gamma=0$. In our case, this corresponds to two different event horizon, given as
 \begin{align}
 r_+=M+\sqrt{M^2-a^2},\\
 r_-=M-\sqrt{M^2-a^2}.
 \end{align}
 For simplicity, we rename $\sqrt{M^2-a^2}$ as $\Delta$.  The black hole is extremal when $\Delta=0$ is satisfied, and the non-extremal case corresponds to $\Delta>0$. $\Delta$ is infinitesimal for both those two cases.

The Hawking temperature $T_{\rm H}$, the Bekenstein-Hawking entropy $S_{\rm BH}$ and the angular
velocity $\Omega_H$ of the black hole are
\begin{equation}\begin{array}{l}
T_{\rm H}=\frac{r_{+}-r_{-}}{4 \pi r_{0} r_{+}}, \\

S_{\rm BH}=\pi r_{0} r_{+}, \\
\Omega_{\rm H}=\frac{a}{r_{0} r_{+}}.
\end{array}
\end{equation}

Considering the special  asymptotical performance of the black hole as well as the linear divergence of the action, its black hole mechanics needs to be considered carefully.
In~\cite{Clement:2002mb} they use the Hamitonian method developed by Brown and York in \cite{Brown:1992br} with a renormalized action,
\begin{equation}\label{10}
\tilde{S}=S_{(g)}+S_{(m)}-S_{(0)},
\end{equation}
where the first two part is the total action given by our classical solution Eq.~(\ref{metric1}) and Eq.~(\ref{fieldeq}), while the third term is given by the corresponding divergent background~(\ref{asympdia}).

In this case
\begin{equation}\label{mass}
\mathcal{M}=\frac{M}{2}.
\end{equation}
\begin{equation}\label{angular}
  J=\frac{a r_{0}}{2}.
\end{equation}
It is the 	fixed charge that make the absence of electric charge coupling with the potential in its first law of black hole.

Eventually, we have
\begin{equation}
d\mathcal{M}=T_{\rm H} d S_{\rm B H}+\Omega_{\rm H} dJ.
\end{equation}
as its first law of black hole thermodynamics. It should be noted that  before coming to a meaningful results we have subtracted the background action by using Hamiton-Jacobi methods.

\section{Komar integral approach to the conserved quantity}\label{komar}

It is well known that the conserved charge for the black hole thermodynamics can be obtained from different methods. Wald have obtained the general results for the asymptotically flat black hole. In this section, we want to deduce the conserved charge in our case using the Komar-type formulae for later simplicity.
To do this, we need to rewrite the metric in a more applicable form.
\begin{equation}
ds^2=-\frac{\Gamma-a^2sin^2\theta}{r_0r}dt^2+\frac{r_0r}{\Gamma}dr^2+r_0rd\Omega-2asin^2\theta r_0rdt d\phi.
\end{equation}
The Komar-type conserved integral is given by(in our notation)
\begin{equation}
\label{Komarintegral}\begin{array}{l}
\mathcal{M}=-\frac{1}{16 \pi}  \int_{S_{\infty}} \nabla^{\alpha} \xi_{(t)}^{\beta} d S_{\alpha \beta}, \\
J=\frac{1}{16 \pi} \int_{S_{\infty}} \nabla^{\alpha} \xi_{(\phi)}^{\beta} d S_{\alpha \beta}.
\end{array}
\end{equation}
The $\xi_{t}+\Omega_H\xi_{\phi}$ is $\partial_{t}+\Omega_H\partial_{\phi}$, the surface element is given as
\begin{equation}
d S_{\alpha \beta}=-2 n_{[\alpha} r_{\beta]} \sqrt{\sigma} d^{2} \theta.
\end{equation}
$S_{\infty}$ is the spatial infinity for chosen 3-hypersurface. The $\sigma$ is the induced metric on $S_{\infty}$, $n$ is the normal vector to our chosen three dimensional spatial hypersurface $t=const$.  The $r$ is the  vector normal to $S_{\infty}$ living in chosen spatial slice of the foliation.
\begin{align}
&n_\alpha=-\sqrt{\frac{r}{r_0}}(1-\frac{M}{r})\partial_{\alpha}t, \\
&r^\alpha=\sqrt{\frac{r}{r_0}}(1-\frac{M}{r})
\frac{\partial x^\alpha}{\partial r}.
\end{align}
Then if one just calculate the integral given in Eq.~(\ref{Komarintegral}), one may find it is linearly divergent with the growing of the radius, while fortunately this term will not be parameter dependent if we set $r_0$ as constant. We can calculate the same integral for the chosen background. One might find that if we subtract the background contribution, it turns out to be
\begin{equation}
\mathcal{M}=\frac{M}{2},
\end{equation}
which gives exactly the same answer as Eq.~(\ref{mass}). Similarly, Eq.~(\ref{angular}) will be recovered from the second line of Eq.~(\ref{Komarintegral}).

From above discussion, one may see that in this linear rotating black hole, the conserved quantity will consist of two parts: one part is parameter dependent and convergent; while another part is parameter independent as well as infinite. The finite part is related to our black hole thermodynamic~\cite{Hawking:1995ap,Clement:2002mb,Wald:1993nt}.

It is amazing that when we adopt Komar formulae, the only things should be incorporated into our case is to subtract the divergent part caused by the background contribution.

\section{classical gedanken experiment}\label{CGE}

There are many ways to conduct the gedanken experiment to test the validity of the WCCC. In this section, we briefly discuss whether we can violate the WCCC with a test particle with a large enough angular momentum.

The lagrangian for a test particle is given by
\begin{equation}
L=\frac{1}{2} m g_{\mu \nu} \frac{d x^{\mu}}{d \tau} \frac{d x^{\nu}}{d \tau}.
\end{equation}
Then the equation of motion for the particle can be derived as
\begin{equation}
\frac{d^{2} x^{\mu}}{d \tau^{2}}+\Gamma_{a \beta}^{\mu} \frac{d x^{a}}{d \tau} \frac{d x^{\beta}}{d \tau}=0.
\end{equation}
The energy and angular momentum are given as
\begin{equation}\begin{array}{l}
\delta E=-P_{t}=-\frac{\partial L}{\partial t}=-m g_{0 \nu} \frac{d x^{\nu}}{d \tau}, \\
\delta J=P_{\phi}=\frac{\partial L}{\partial \dot{\phi}}=m g_{3 \nu} \frac{d x^{\nu}}{d \tau}.
\end{array}
\end{equation}
The 4-velocity of a massive particle satisfies
\begin{equation}g_{\mu \nu} \frac{d x^{\mu}}{d \tau} \frac{d x^{\nu}}{d \tau}=\frac{1}{m^{2}} g^{\mu \nu} P_{\mu} P_{\nu}=-1.\end{equation}
Combining above expression, we have
\begin{equation}\begin{aligned}
\delta E=& \frac{g^{03}}{g^{00}} \delta J-\frac{1}{g^{00}}\left[\left(g^{03}\right)^{2} \delta J^{2}-g^{00} g^{33} \delta J^{2}\right.\\
&\left.-g^{00}\left(g^{11} P_{r}^{2}+g^{22} P_{\theta}^{2}+m^{2}\right)\right]^{\frac{1}{2}},
\end{aligned}
\end{equation}
where we have used the future directed condition $dt/d\tau >0$, that is
 \begin{equation}
 \delta E>-\frac{g_{03}}{g_{33}} \delta J.
 \end{equation}

Hence, for perturbation processes, the energy $\delta E$ and angular momentum $\delta J$ must satisfy
 \begin{equation}\label{i1}
\delta E>\Omega_{\mathrm{H}} \delta J.
\end{equation}
On the other hand, if we want to overspin the black hole, we need condition
\begin{equation}\label{i2}
M+\delta E<a+\delta a.
\end{equation}
For the extremal black hole($M=a$), Eq.(\ref{i1}) and Eq.(\ref{i2}) can be rewritten as
\begin{align}
\delta J<r_0\delta E,\\
\delta J>r_0\delta E,
\end{align}
which cannot be satisfied simultaneously as Wald's foundation work suggested.

If we consider the near-extremal black hole, that might not be the same case as extremal. To do this, we rewrite Eq.~(\ref{i1}) and Eq.~(\ref{i2}) as
\begin{align}
&\delta J<\frac{r_0^2( E+\Delta)\delta E}{J},\\
&\delta J>r_0E-J+r_0\delta E.
\end{align}
For the near-extremal case, it is true $\mathcal{M}>\frac{J}{r_0}$, which results $\frac{r_0^2\mathcal{M}}{J}>r_0$. Certainly the values satisfying these two inequality simultaneously exist. Thus, the event horizon of the near-extremal black hole can be destroyed and the weak cosmic censorship conjecture can be violated.

\section{Review of Wald's geometrical formulation}\label{W1}
In this section, we discuss the Wald formulation for a linear dilaton black hole. We review the steps taken in~\cite{Wald:1993nt,Iyer:1994ys,Iyer:1995kg,Wald:1999wa,SoWa17}.

For simplicity, we note that we denote the variation as
\begin{equation}
\delta \phi=\frac{d\phi}{d\lambda}|_{\lambda=0},
\end{equation}
\begin{equation}
\delta^2 \phi=\frac{d^2\phi}{d\lambda^2}|_{\lambda=0}.
\end{equation}

The off-shell variation is given as
\begin{equation}\label{variation1}
\delta L=E_{\phi}\delta\phi+d\Theta
.\end{equation}
$E_{\phi}$ corresponds to the equation of motion, while $\Theta=\Theta(\phi,\delta\phi)$ defines a 3-form.
If we substitute $\delta\phi=\mathcal{L}_{\xi}\phi$ into the variation (where $\xi$ is a killing vector defined by the metric), with the well-known identity $\mathcal{L}_{\xi}=di_{\xi}+i_{\xi}d$, where $i_{\xi}$ is the interior product of
the differential form,
$d$ is the differential operator. Eq.(\ref{variation1}) is then
\begin{equation}
di_{\xi}L=E_{\phi}\delta\phi+d\Theta.
\end{equation}
If we take  $E_{\phi}=0$, then we have
\begin{equation}\label{current1}
d(i_{\xi}L-\Theta)=0.
\end{equation}

We rewrite
\begin{equation}
\label{current2}
i_{\xi}L-\Theta=-J_{\xi}.
\end{equation}
It is easy to see that the $J_{\xi}$ is a closed 3-form defined by $\xi$ iff the equation of motion is satisfied. According to the Poincare lemma, it is locally exact.
Wald further showed that~\cite{Iyer:1995kg}
\begin{equation}
\label{current3}
J_{\xi}=C_{\xi}+d Q_{\xi},
\end{equation}
where $C_{\xi}$ and $Q_{\xi}$ denote the constraints and noether charge 2-form, respectively.
If we further assume that $\xi$ is invariant with respect to the variation, then take another variation of $J_{\xi}$. Combine Eq.~(\ref{current1}), Eq.~(\ref{current2}) and Eq.~(\ref{current3}) and subtract the same term with inverse order of variation to obtain
\begin{equation}\label{differential1}
d\left[\delta Q_{\xi}-\xi \cdot \Theta(\phi, \delta \phi)\right]=\omega\left(\phi, \delta \phi, \mathcal{L}_{\xi} \phi\right)-\xi \cdot E_{\phi} \delta \phi-\delta C_{\xi},
\end{equation}
where we identify
\begin{equation}
\omega\left(\phi, \delta_{1} \phi, \delta_{2} \phi\right)=\delta_{1} \Theta\left(\phi, \delta_{2} \phi\right)-\delta_{2} \Theta\left(\phi, \delta_{1} \phi\right).
\end{equation}
Then one may further obtain from Eq.~(\ref{differential1}) higher-rank variation equality

\begin{equation}\label{differential2}
d\left[\delta^2Q_{\xi}-\xi \cdot \delta\Theta(\phi, \delta \phi)\right]=\omega\left(\phi, \delta \phi, \mathcal{L}_{\xi} \delta\phi\right)-\xi \cdot\delta E_{\phi} \delta \phi-\delta^2C_{\xi}.
\end{equation}

Even further like in~\cite{Wang:2020vpn}, but in this article, what will be useful is only the first two order.

In Wald's paper~\cite{Wald:1999wa}, the corresponding conserved quantity for the black hole in the asymptotically-flat case is given as
\begin{equation}\label{old}
\begin{array}{c}
\mathcal{E}=\int_{\infty}(Q[t]-t \cdot \Theta(\phi, \delta \phi)), \\
\mathcal{J}=-\int_{\infty} Q[\varphi],
\end{array}
\end{equation}
where $\xi^{a}=t^{a}+\Omega_{\mathcal{H}} \varphi^{a}$ is a time-like killing vector field of a stationary black hole as mentioned above.

Eq.(\ref{old}) is not the case in our discussion. To use the right hand side of the above equation correctly, for linear dilaton black hole, we need to subtract the corresponding divergence from the conserved quantity.
Following with the discussion in Sec.~\ref{bh}, \ref{komar} and \ref{CGE}, we may define
\begin{equation}\begin{array}{c}
\mathcal{E}+\mathcal{E}_0=\int_{\infty}(Q[t]-t \cdot \Theta(\phi, \delta \phi)), \\
\mathcal{J}=-\int_{\infty} Q[\varphi].
\end{array}
\end{equation}
The $\mathcal{E}_0$ is the divergent part from linear dilaton background, and $\mathcal{E}$ is the finite part that contributes to the first law of thermodynamics hence we may rewrite this as
\begin{equation}\begin{array}{c}
\mathcal{E}=\int_{\infty}(Q[t]-t \cdot \Theta(\phi, \delta \phi))-\mathcal{E}_0, \\
\mathcal{J}=-\int_{\infty} Q[\varphi].
\end{array}
\end{equation}
For more general case, we may generalise the Eq.~(\ref{current3}) and Eq.~(20) in~\cite{Wald:1999wa} with
\begin{equation}\begin{array}{c}
\mathcal{E}+\mathcal{E}_0=\int_{\infty}(Q[t]-t \cdot \Theta(\phi, \delta \phi)), \\
\mathcal{J}=-\int_{\infty} Q[\varphi].
\end{array}
\end{equation}
Here the $\mathcal{E}_0$ and $\mathcal{J}_0$ describe the quantity defined by what we choose as background. In general, $\mathcal{J}_0=0$ accounts for non-rotating background.

Using results from above discussion, we can rewrite Eq.~(\ref{differential1}) and Eq.~(\ref{differential2}) in its integral form. (We choose the domain as $\Sigma=\mathcal{H}\cup\Sigma_0$, these 3-surface is bounded by the bifurcate surface and $S_{\infty}$. )
\begin{equation}\label{integral1}\begin{aligned}
\int_{\partial \Sigma}\left[\delta Q_{X}-\iota_{X} \Theta(\phi, \delta \phi)\right]=& \int_{\Sigma} \omega\left(\phi ; \delta \phi, \mathcal{L}_{X} \phi\right)-\int_{\Sigma} \delta C_{X} \\
&-\int_{\Sigma} l_{X}(E(\phi) \cdot \delta \phi).
\end{aligned}
\end{equation}
\begin{equation}\label{integral2}\begin{aligned}
\mathcal{E}_{\Sigma}(\phi ; \delta \phi)=& \int_{\partial \Sigma}\left[\delta^{2} Q_{\xi}-\imath_{\xi} \delta \Theta(\phi, \delta \phi)(\phi, \delta \phi)\right]+\int_{\Sigma} \delta^{2} C_{\xi} \\
&+\int_{\Sigma} l_{\xi}(\delta E \cdot \delta \phi),
\end{aligned}
\end{equation}
where
\begin{equation}
\mathcal{E}_{\Sigma}(\phi ; \delta \phi) \equiv \int_{\Sigma} \omega\left(\phi ; \delta \phi, \mathcal{L}_{\xi} \delta \phi\right).
\end{equation}
Stokes theorem has been used in between.
The terms $\int_{\infty}\left[\delta Q_{X}-\iota_{X} \Theta(\phi, \delta \phi)(\phi, \delta \phi)\right]$ for the asymptotically flat case with suitable choice of $\xi$ account  for $\delta \mathcal{E}-\Omega_{H}^{(\mu)} \delta \mathcal{J}_{(\mu)}$ exactly.

\section{variational Inequality}\label{v1}
In this section, we will obtain the inequality in EMDA theory required to discuss weak cosmic censorship conjecture. As shown before, the variational inequality is easy to be obtained from the lagrangian 4-from with the assistance of Lie derivative, so we start from the lagrangian description of the theory discussed in Sec.~\ref{bh}.
\begin{align}
L=&\nonumber\frac{\tilde{\epsilon}}{16\pi}[R-2\partial_{\mu}\phi\partial^{\mu}\phi-\frac{e^{4\phi}}{2}\partial_{\mu}\kappa\partial^{\mu}\kappa-e^{-2\phi}F_{\mu\nu}F^{\mu\nu}\\&-\kappa F_{\mu\nu}\star F^{\mu\nu}].
\end{align}
We may divide it into two parts, the gravitational part and the matter parts
\begin{equation}
L= \frac{\tilde{\epsilon}}{16\pi}R+L_{others}.
\end{equation}
As we introduce the extra matter source (perturbation) term $T_{ab}(\lambda)$ with $T_{ab}(0)=0$  into the the system, It turns out that the equation of motion and $\Theta$ can be given like
\begin{equation}\begin{aligned}
&R_{ab}-\frac{1}{2}Rg_{ab}=8\pi (T_{ab}^{DIL}+T^{EM}_{ab}+T^{axion}_{ab}+T_{ab}),
\end{aligned}\end{equation}
\begin{equation}\begin{array}{c}
\nabla^{\mu} \nabla_{\mu} \phi=\frac{1}{2} \mathrm{e}^{-2 \Phi} F^{2}+\frac{1}{2} \mathrm{e}^{4 \Phi}(\partial a)^{2} ,\\
\nabla_{\mu}\left(\mathrm{e}^{-2 \Phi} F^{\mu \nu}+\kappa\star F^{\mu \nu}\right)=4 \pi j^\nu,\\
\nabla_{\mu}\left(\mathrm{e}^{4 \Phi} g^{\mu \nu} \partial_{\nu} a\right)+(\star F)_{\mu \nu} F^{\mu \nu}=0.
\end{array}
\end{equation}
\begin{equation}
\Theta(\phi, \delta \phi) =\Theta^{\mathrm{GR}}(\phi, \delta \phi)+\Theta^{\mathrm{Matter}}(\phi, \delta \phi).
\end{equation}
For latter convenience, we may divide the $\Theta^{\mathrm{Matter}}(\phi, \delta \phi)$ as
\begin{equation}
\Theta^{\mathrm{Matter}}(\phi, \delta \phi)=\Theta^{\mathrm{1}}(\phi, \delta \phi)+\Theta^{\mathrm{CS}}(\phi, \delta \phi),
\end{equation}
where $\Theta^{\mathrm{CS}}(\phi, \delta \phi)$ comes from Chern-Simons (CS) part, and the remaining stuff is named as $\Theta^{\mathrm{1}}(\phi, \delta \phi)$.
\begin{eqnarray}
 \Theta_{a b c}^{\mathrm{GR}}(\phi, \delta \phi) &=& \frac{1}{16 \pi} \epsilon_{d a b c} g^{d e} g^{f g}\left(\nabla_{g} \delta g_{e f}-\nabla_{e} \delta g_{f g}\right) , \\\nonumber
  \Theta_{a b c}^{\mathrm{1}}(\phi, \delta \phi)&=& -\frac{1}{4 \pi} \epsilon_{d a b c}e^{-2\phi} F^{d e} \delta A_{e}
  \\ \nonumber&-&\frac{1}{4 \pi} \epsilon_{d a b c}\left(\nabla^{d} \phi\right) \delta \phi
\\&-&\frac{1}{16\pi} \epsilon_{d a b c}e^{4\phi}\left(\nabla^{d} \kappa\right) \delta \kappa, \\
  \Theta_{a b c}^{\mathrm{CS}}(\phi, \delta \phi) &=&-\frac{1}{8\pi}\kappa F_{[ab} \delta A_{c]}.
\end{eqnarray}
\begin{eqnarray}
   \omega_{a b c}^{\mathrm{GR}}&=& \frac{1}{16 \pi} \epsilon_{d a b c} w^{d} , \\
  \omega_{a b c}^{\mathrm{1}}&=& \delta_1\Theta_{a b c}^{\mathrm{1}}(\phi, \delta_2 \phi)-\delta_2\Theta_{a b c}^{\mathrm{1}}(\phi, \delta_1 \phi), \\
  \omega_{a b c}^{\mathrm{CS}} &=& -\delta_1(\frac{1}{8 \pi}\kappa F_{[ab} \delta_2 A_{c]})+\delta_2(\frac{1}{8 \pi}\kappa F_{[ab} \delta_1 A_{c]}).
\end{eqnarray}
\begin{equation}w^{a}=P^{abcdef}\left(\delta_{2} g_{b c} \nabla_{d} \delta_{1} g_{e f}-\delta_{1} g_{b c} \nabla_{d} \delta_{2} g_{e f}\right).
\end{equation}
\begin{equation}
\begin{aligned}
P^{abcdef}=& g^{a e} g^{f b} g^{c d}-\frac{1}{2} g^{a d} g^{b e} g^{f c} -\frac{1}{2} g^{a b} g^{c d} g^{e f}\\&-\frac{1}{2} g^{b c} g^{a e} g^{f d}+\frac{1}{2} g^{b c} g^{a d} g^{e f}.
\end{aligned}
\end{equation}
Hence the constraint will be like
\begin{equation}
C_{a b c d}=\epsilon_{e b c d}\left(T_{a}^{e}+A_{a} j^{e}\right).
\end{equation}
Noether charge is
\begin{equation}
Q_{\xi}=Q_{\xi}^{\mathrm{GR}}+Q_{\xi}^{\mathrm{EM}}+Q_{\xi}^{\mathrm{CS}},
\end{equation}
where
\begin{eqnarray}
  \left(Q_{\xi}\right)^{\mathrm{GR}}_{a b} &=& -\frac{1}{16 \pi} \epsilon_{a b c d} \nabla^{c} \xi^{d}, \\
  \left(Q_{\xi}\right)_{a b}^{\mathrm{CS}} &=& -\frac{1}{8 \pi} \kappa F_{ab}\xi^e A_e, \\
  \left(Q_{\xi}^{1}\right)_{a b} &=& -\frac{1}{8\pi} \epsilon_{a b c d} e^{-2\phi} F^{cd} A_{e} \xi^{e}.
\end{eqnarray}
%
Following the set-up in \cite{Jiang:2019vww,Gao:2001ut}, we choose $\Sigma=\Sigma_0 \cup \mathcal{H}$. The 3-hypersurface starts from the bifurcate surface where no collision occurs, to its future after which the collision has occurred, then extends spatially to spatial infinity. From this we know that $\Sigma$ is bounded by a bifurcate surface noted as $B$, and spatial infinity $S_\infty$.

We assume the stability of the non-extremal black hole, which means that it will evolve into the same black hole with different parameters. With all these set-ups, one may rewrite Eq.~(\ref{integral1}) as
\begin{equation}
 -\int_{B}[\delta Q_X-i_\xi\Theta(\phi,\delta \phi)] +\int_{\infty}[\delta Q _X-i_\xi\Theta(\phi,\delta \phi)]=-\int_{\Sigma}C_\xi
  .
\end{equation}
  Where $\omega=0$ because that $\xi$ is the killing vector field as mentioned before, the equation of motion is satisfied so that the last term vanishes. The first term vanishes due to there is no perturbation near the bifurcate surface $B$ till the very late time.
  Then It turns out to be
  \begin{equation}
  \delta \mathcal{M}-\Omega_{\rm H} \delta J\geq 0,
  \end{equation}
  where the Null Energy Condition has been used~\cite{Gao:2001ut}. It's worth noting that this result exactly corresponds to what we obtain in section~\ref{CGE}.

One thing should be noted is that although we adopt a third method to calculate the conserved charge of the black hole, we come to the same variation expression for charge as with the BY and Komar integral formulae.

With $\mathcal{E}_{\Sigma}=\mathcal{E}_{\mathcal{H}}+\mathcal{E}_{\Sigma_0}$, Eq.~(\ref{integral2})  can be represented as

\begin{equation}\begin{aligned}
\mathcal{E}_{\mathcal{H}}+\mathcal{E}_{\Sigma_0}=& \int_{\partial \Sigma}\left[\delta^{2} Q_{\xi}-\imath_{\xi} \delta \Theta(\phi, \delta \phi)\right]+\int_{\Sigma} \delta^{2} C_{\xi} \\
&+\int_{\Sigma} l_{\xi}(\delta E \cdot \delta \phi).
\end{aligned}
\end{equation}
The first term satisfies~\cite{Jiang:2019vww}
\begin{equation}
\mathcal{E}_{\mathcal{H}}\geq \mathcal{E}_{\mathcal{H}}^{C S}.
\end{equation}
Then we have
\begin{equation}
\begin{aligned}
8 \pi \mathcal{E}_{\mathcal{H}}^{C S} &=-\int_{\mathcal{H}} \delta_{1}\left(\kappa F_{[a b} \mathcal{L}_{\xi} \delta A_{c]}\right)+\int_{\mathcal{H}} \mathcal{L}_{\xi} \delta\left(\kappa F_{[a b} \delta_{1} A_{c]}\right), \\
&=\int_{\mathcal{H}} \mathcal{L}_{\xi} \delta \kappa F_{[a b} \delta A_{c]}+\int_{\mathcal{H}} \kappa \mathcal{L}_{\xi} \delta F_{[a b} \delta A_{c]} \\
&-\int_{\mathcal{H}} \delta \kappa F_{[a b} \mathcal{L}_{\xi} \delta A_{c]}+\int_{\mathcal{H}} \kappa \delta F_{[a b} \mathcal{L}_{\xi} \delta A_{c]}
.\end{aligned}
\end{equation}
With the consideration that $\xi^a=0$ at the bifurcate surface, as well as the gauge condition
 $\xi^a\delta A_a=0$, the right hand side is
\begin{equation}
8 \pi \mathcal{E}_{\mathcal{H}}^{C S}=\int_{\mathcal{H}} d\left(\xi \cdot\left(\kappa \delta F_{[a b} \delta A_{c]}\right)\right)=0.
\end{equation}

The second term can be obtained by reusing Eq.~(\ref{integral2}) on $\Sigma_0$.
\begin{equation}
\begin{aligned}
\mathcal{E}_{\Sigma_0}(\phi ; \delta \phi)=& \int_{\partial \Sigma_0}\left[\delta^{2} Q_{\xi}-\imath_{\xi} \delta \Theta(\phi, \delta \phi)\right]+\int_{\Sigma_0} \delta^{2} C_{\xi} \\
&+\int_{\Sigma_0} l_{\xi}(\delta E \cdot \delta \phi),
\end{aligned}
\end{equation}
which can be reduced to~\cite{SoWa17}
\begin{equation}
\mathcal{E}_{\Sigma_0}(\phi ; \delta \phi)=-T_{\rm H} \delta^{2} S_{\rm B H}.
\end{equation}

Considering the null energy condition, our second order inequality can be written as~\cite{SoWa17}
\begin{equation}
  \delta^{2} \mathcal{M}+T_{\rm H} \delta^{2} S_{\rm B H}-\Omega_{\rm H} \delta^{2} J \geq 0.
\end{equation}
One may refer to~\cite{SoWa17,Jiang:2019vww} for more details.

\section{Gedenken experiment}\label{t1}

From Eq.~(\ref{GAMMA}), We may define a function as
\begin{equation}
j(\lambda)=M^2-a^2.
\end{equation}
It is worth noting that the M appearing here does not corresponding to $\mathcal{M}$ because of the Eq.~(\ref{mass}).
Then we may write
\begin{align}
\label{secondorderj}
j(\lambda)=\nonumber&M^2-a^2+(2M\delta M-2a\delta a)\lambda+\\
&(\delta M^2- \delta a^2+M\delta^2M-a\delta^2a)\lambda^2.
\end{align}
Actually, one may consider more complicated case involving higher rank variation here, but one still need higher rank inequality to evaluate our $j(\lambda)$.  Here what we need is just to consider whether if this $j(\lambda)\ge0$ still be held in our case. We refer to inequality obtained for help.
\begin{equation}\label{inequ1}
\frac{1}{2}\delta M-\Omega_{\rm H} \delta J\geq 0.
\end{equation}

\begin{equation}\label{inequ2}
\frac{1}{2}\delta^2 M-\Omega_{\rm H} \delta^2 J\geq -T_{\rm H} \delta^2 S_{\rm B H}.
\end{equation}
Following the same setting as in~\cite{SoWa17},We rewrite

\begin{equation}
\delta^2 r_{+BH}= \frac{-1}{\Delta^3}(M\delta M-a\delta a)^2+\frac{1}{\Delta}(\delta M^2-\delta a^2).
\end{equation}
We may reexpress Eq.~(\ref{inequ1}) and Eq.~(\ref{inequ2}) as
\begin{equation}
M\delta M-a\delta a\geq -\Delta\delta M.
\end{equation}
\begin{equation}
\delta^2M-\frac{a\delta^2a}{r_+}\geq \frac{(M\delta M-a\delta a)^2}{\Delta^2 r_+}-\frac{(\delta M)^2-(\delta a)^2}{r_+}.
\end{equation}
If we substitute the first inequality into the second one, then
\begin{equation}
\delta^2M-\frac{a\delta^2a}{r_+}\geq \frac{(\delta M)^2}{r_+}-\frac{(\delta M)^2-(\delta a)^2}{r_+}.
\end{equation}
Ignoring the $\Delta$ term, we have

\begin{equation}
\delta M^2- \delta a^2+M\delta^2M-a\delta^2a\geq(\delta M)^2.
\end{equation}

With this for help, we may firstly assess $j(\lambda)$ to its first order, that is to say, we only consider the $\lambda$-term in Eq.~(\ref{secondorderj}) and  Eq.~(\ref{inequ1}).
Then we will automatically receive the same results as in Sec.~\ref{CGE}. The near extremal case may not respect the WCCC in our case to first order. But for the extremal case $(\Delta=0)$, no violation occurs as discussed in Sec.~\ref{komar}.

If we consider  Eq.~(\ref{inequ2}) and $j(\lambda)$ to its second order, then we will obtain
\begin{equation}
j(\lambda)\geq(\Delta-\delta M\lambda)^2\geq 0.
\end{equation}

As expected, the WCCC is restored for the near-extremal black hole if we consider the modification up to second order approximation.

\section{Conclusion}\label{conclusion}

In this paper, we use Iyer-Wald formalism to extract the invariant information of the linear dilaton black hole~\cite{Clement:2002mb}. To confirm the conclusion is consistent with one might obtain by using other methods in this case, we compare the vaiation of conserved charge obtained among several methods. The calculation shows consistency among these methods. Then we use two different methods to discuss the WCCC for a rotating linear dilaton black hole in the EMDA theory. Up to first order approximation, we can come to the same conclusion for both extremal and near-extremal black hole via these two methods as Wald said in~\cite{SoWa17}. WCCC is well preserved for extremal black hole, but not so for near-extremal black hole. To second order, the WCCC is preserved in our case as expected  for both extremal and nearly extremal black hole, which implies that if we test the nearly extremal black hole with the test particle up to second order precision, WCCC is well protected.

\section*{Acknowledgement}

We thank Jie Jiang for enlightening discussions and many useful suggestions. This work was supported by the basic scientific research business expenses of the central university and the Open Project of Key Laboratory for Magnetism and Magnetic Materials of the Ministry of Education, Lanzhou University (LZUMMM2020010). Fei Qu acknowledges support from the Fundamental Research Funds for the Central Universities (Grants No. lzujbky-2020-it04). Si-Jiang Yang acknowledges support from National Natural Science Foundation of China (Grants No. 11875151).

\bibliography{WeakCosmic}
\bibliographystyle{apsrev4-1}

\end{document}